\documentclass[aps,prl,reprint,superscriptaddress,nofootinbib]{revtex4-2}

\usepackage{amsmath,amssymb,bm}
\usepackage{graphicx}
\usepackage{hyperref}
\usepackage{xcolor}
\usepackage{float}
\newcommand{\fnl}{f_{\rm NL}}
\newcommand{\dd}{\delta}
\newcommand{\kk}{\bm{k}}
\newcommand{\xx}{\bm{x}}

\newcommand{\web}{\mathcal{W}}
\newcommand{\Tr}{\mathrm{Tr}}

\begin{document}

\title{
Primordial non-Gaussianity as a large-scale cosmic-web bias
}

\author{Kilian Peraza Vera}
\affiliation{Instituto de Astrofísica de Canarias, Vía Láctea s/n, E-38205, La Laguna, Tenerife, Spain}
\affiliation{Departamento de Astrofísica, Universidad de La Laguna, E-38206, La Laguna, Tenerife, Spain}

\author{Francisco-Shu Kitaura}
\email{fkitaura@ull.edu.es}
\affiliation{Instituto de Astrofísica de Canarias, Vía Láctea s/n, E-38205, La Laguna, Tenerife, Spain}
\affiliation{Departamento de Astrofísica, Universidad de La Laguna, E-38206, La Laguna, Tenerife, Spain}

\author{Francesco Sinigaglia}
\affiliation{Institute for Fundamental Physics of the Universe, Via Beirut 2, I-34151 Trieste, Italy}
\affiliation{SISSA -- International School for Advanced Studies, Via Bonomea 265, I-34136 Trieste, Italy}
\affiliation{INAF -- Osservatorio Astronomico di Trieste, Via G. B. Tiepolo 11, I-34131 Trieste, Italy}
\affiliation{INFN -- National Institute for Nuclear Physics, Via Valerio 2, I-34127 Trieste, Italy}

\date{\today}

\begin{abstract}
Primordial non-Gaussianity is a unique probe of the physics of inflation,
but its large-scale-structure signature is difficult to measure because it
appears on the largest, most cosmic-variance-limited modes.  The standard
view of local-type primordial non-Gaussianity is isotropic: it induces a
scale-dependent correction to tracer bias,
$\Delta b(k)\propto \fnl k^{-2}$, on large scales.  Here we show that the
same large-scale response can be reformulated as an anisotropic, nonlocal
environmental-bias effect.  Motivated by the spectral cosmic-web hierarchy
of Kitaura and Sinigaglia, we define a Hessian field
\begin{equation}
\mathcal{H}_{ij}(\xx)
\equiv
\partial_i\partial_j\nabla^{-4}\dd(\xx),
\end{equation}
whose trace obeys $\Tr \mathcal{H}=\nabla^{-2}\dd$ and therefore carries
the same $k^{-2}$ enhancement that appears in the Dalal et al.
scale-dependent-bias term.  Diagonalizing $\mathcal{H}_{ij}$ defines four
large-scale web environments, analogous to knots, filaments, sheets, and
voids, although operating on scales much larger than the conventional
nonlinear cosmic web.  We assign one weight to each environment and test the construction with
the \textsc{AbacusPNG} suite of local-PNG $N$-body simulations, using
$f_{\rm NL}=\{-100,-30,30,100\}$.  We show that these four weights
reproduce the large-scale $k^{-2}$ rise of the tracer bias.
We further show that existing halo catalogs without PNG can be post-processed to
imprint an effective $\fnl$ signature.  The fitted weights exhibit an
ordered hierarchy: for positive $\fnl$, knot-like regions receive the
largest weights, followed by filaments, sheets, and voids, while the order
reverses for negative $\fnl$.  Extending the bias basis with the standard
tidal-field tensor,
$\mathcal{T}_{ij}\equiv \partial_i\partial_j\nabla^{-2}\delta$, and
assigning four additional environment weights to its knot-, filament-,
sheet-, and void-like regions improves the reconstruction toward higher
wavenumbers, where the pure $\nabla^{-4}$ Hessian response alone becomes
insufficient.  This establishes local PNG as a large-scale nonlocal
cosmic-web bias. {Beyond mock generation, the resulting
cosmic-web response motivates an observational statistic based on changes
in tracer occupation of large-scale environments relative to a matched
Gaussian reference, complementary to the conventional low-$k$ power spectrum.}
\end{abstract}

\maketitle

\paragraph{Introduction.}
The statistics of primordial fluctuations encode information about the
dynamics of inflation \cite{2004PhR...402..103B,2010AdAst2010E..72C}.
In the simplest single-field slow-roll scenarios, primordial fluctuations
are close to Gaussian, and the squeezed-limit bispectrum is suppressed by
a consistency relation
\cite{2003JHEP...05..013M,2003NuPhB.667..119A,2004JCAP...10..006C}.
Conversely, a detection of local-type primordial non-Gaussianity,
currently tightly constrained by the CMB but still a major target for
large-scale-structure surveys
\cite{2020A&A...641A...9P,2010AdAst2010E..76B},
\begin{equation}
\Phi(\xx)
=
\phi(\xx)
+
\fnl
\left[
\phi^2(\xx)-\langle \phi^2\rangle
\right],
\label{eq:localpng}
\end{equation}
at a level significantly above the single-field expectation would point to
additional fields, nontrivial dynamics, or nonstandard initial conditions.

Large-scale structure offers an especially promising probe because local
PNG couples long- and short-wavelength modes.  This coupling changes the
abundance and clustering of biased tracers, producing the well-known
scale-dependent bias
\cite{2008PhRvD..77l3514D,2008JCAP...08..031S,2008ApJ...677L..77M,2010PhRvD..81b3006D},
which motivates power-spectrum, multitracer, correlation-function, and
bispectrum strategies in current and forthcoming surveys
\cite{2011PhRvD..84h3509H,2018MNRAS.478.1341K,2019MNRAS.489.1950B,2023MNRAS.523..603R,2024PhRvD.109l3511H}
\begin{equation}
\Delta b_{\rm PNG}(k,z)
=
\fnl\, b_\Phi\, \mathcal{M}^{-1}(k,z),
\label{eq:sdb}
\end{equation}
where
\begin{equation}
\delta(\kk,z)=\mathcal{M}(k,z)\Phi(\kk),
\qquad
\mathcal{M}(k,z)
=
\frac{2}{3}
\frac{k^2 T(k)D(z)}
{\Omega_m H_0^2}.
\end{equation}
On sufficiently large scales, $T(k)\rightarrow 1$, and therefore
$\Delta b_{\rm PNG}\propto \fnl k^{-2}$.  This signature is powerful but
challenging: it lives on the largest modes, where finite-volume
fluctuations, survey systematics, and relativistic projection effects are
most severe. {A measurement based on environmental
number-density contrasts offers a different compression of the survey:
rather than isolating the PNG response through a small set of low-$k$
band powers, it compares tracer occupation across spatially distributed
regions. The impact of survey selection can then be modeled and tested
through environment-conditioned counts and matched random catalogues.
This does not render the web insensitive to contaminated long-wavelength
modes, but the resulting statistic need not respond to systematics in the
same way as the conventional power spectrum; the two are potentially
complementary and can be combined.}

Here we propose a complementary interpretation.  Rather than viewing
Eq.~\eqref{eq:sdb} purely as an isotropic correction to the tracer bias, we
recast it as a large-scale nonlocal environmental-bias effect.  The key
observation is that the operator
\begin{equation}
\Theta(\xx)
\equiv
\Tr\left[
\partial_i\partial_j\nabla^{-4}\delta(\xx)
\right]
=
\nabla^{-2}\delta(\xx)
\label{eq:theta}
\end{equation}
has Fourier-space kernel
\begin{equation}
\Theta(\kk) = -\frac{\delta(\kk)}{k^2},
\label{eq:thetafourier}
\end{equation}
up to sign convention.  Thus, relative to the density field, $\Theta$
carries precisely the large-scale $k^{-2}$ enhancement characteristic of
local PNG bias. {The underlying local-PNG ensemble remains
statistically isotropic; the Hessian eigenvectors and environment classes
provide a spatially resolved geometrical representation of its nonlocal
bias response, rather than a preferred primordial direction.}

\paragraph{Large-scale web operator.}
Following the spectral cosmic-web hierarchy introduced in
Ref.~\cite{2026arXiv260315834K}, we define the symmetric tensor
\begin{equation}
\mathcal{H}_{ij}(\xx;R)
=
\partial_i\partial_j\nabla^{-4}\delta_R(\xx),
\label{eq:hessian}
\end{equation}
where $\delta_R$ is the density field smoothed on a large scale $R$.  The
tensor $\mathcal{H}_{ij}$ is a Hessian-like nonlocal object.  Its
eigenvalues,
\begin{equation}
\lambda_1(\xx)\geq \lambda_2(\xx)\geq \lambda_3(\xx),
\end{equation}
define a four-class partition of space.  In analogy with standard
cosmic-web classifications, we label cells as knot-like, filament-like,
sheet-like, or void-like according to the number of eigenvalues above a
threshold $\lambda_{\rm th}$:
\begin{equation}
\web_H(\xx)
=
\begin{cases}
K, & N_\lambda=3,\\
F, & N_\lambda=2,\\
S, & N_\lambda=1,\\
V, & N_\lambda=0,
\end{cases}
\qquad
N_\lambda
=
\sum_{i=1}^3
\Theta_{\rm H}(\lambda_i-\lambda_{\rm th}),
\label{eq:webclasses}
\end{equation}
where $\Theta_{\rm H}$ is the Heaviside step function.  Because
$\mathcal{H}_{ij}$ is constructed from $\nabla^{-4}\delta$, these
environments are not the usual nonlinear cosmic-web structures.  They are
instead large-scale, nonlocal environments whose trace reproduces the
$k^{-2}$ scaling relevant for local PNG.

To extend the reconstruction beyond the largest scales, we also include
the standard tidal-field tensor
\begin{equation}
\mathcal{T}_{ij}(\xx;R)
=
\partial_i\partial_j\nabla^{-2}\delta_R(\xx),
\label{eq:tidal}
\end{equation}
whose trace is the smoothed density field,
$\Tr\mathcal{T}=\delta_R$.  In Fourier space,
\begin{equation}
\mathcal{T}_{ij}(\kk;R)
=
-\frac{k_i k_j}{k^2}\,\delta_R(\kk),
\label{eq:tidalfourier}
\end{equation}
up to the same sign convention as in Eq.~\eqref{eq:thetafourier}.  The
tidal tensor therefore encodes the anisotropic large-scale environment
without the additional $k^{-2}$ enhancement carried by
$\mathcal{H}_{ij}$.  The two tensors probe complementary contributions:
$\mathcal{H}_{ij}$ captures the nonlocal PNG-like modulation, while
$\mathcal{T}_{ij}$ captures residual tidal dependence that becomes
important toward higher wavenumbers.

We classify $\mathcal{T}_{ij}$ into the same four environments,
\begin{equation}
\web_T(\xx)\in \{K,F,S,V\},
\end{equation}
using the number of tidal eigenvalues above a threshold.  This introduces
a second set of four weights,
\begin{equation}
\bm{v}=(v_K,v_F,v_S,v_V),
\end{equation}
analogous to the Hessian weights $\bm{w}$ but associated with the
$-k^{-2}\delta$ tidal operator rather than with the $-k^{-4}\delta$
Hessian operator.

\paragraph{Environmental weighting.}
The Hessian-weighted tracer field is
\begin{equation}
n_w(\xx)
=
n(\xx)
\sum_{\alpha\in{K,F,S,V}}
w_\alpha\,
\mathbf{1}^{H}_\alpha(\xx),
\label{eq:weightedfield_hessian}
\end{equation}
where $\mathbf{1}^{H}_\alpha$ is the indicator field of the environment
$\alpha$ defined by $\mathcal{H}_{ij}$. In other words,
\[
\mathbf{1}^{H}_{\alpha}(\mathbf{x})
=
\begin{cases}
1, & \text{if } \mathbf{x} \text{ belongs to } \alpha
\text{ according to } \mathcal{H}_{ij},\\
0, & \text{otherwise}.
\end{cases}
\] The corresponding tidal-field
extension is
\begin{eqnarray}
\lefteqn{n_{wv}(\xx)
=}\\&&\hspace{-0.3cm}
n(\xx)
\left[
\sum_{\alpha\in{K,F,S,V}}
w_\alpha\,
\mathbf{1}^{H}_\alpha(\xx)
\right]
\left[
\sum_{\beta\in{K,F,S,V}}
v_\beta\,
\mathbf{1}^{T}_\beta(\xx)
\right], \nonumber
\label{eq:weightedfield_hessian_tidal}
\end{eqnarray}
where $\mathbf{1}^{T}_\beta$ denotes the indicator field of the
environment $\beta$ defined by $\mathcal{T}_{ij}$.  This factorized form
keeps the model compact: instead of fitting sixteen joint Hessian--tidal
weights \citep[][]{2024JCAP...07..083C}, we use only four Hessian weights and four tidal weights.

For each classification we also measure the four environment number
densities,
\begin{eqnarray}
\bar n^H_\alpha
&=&
\frac{1}{V}
\int_V d^3x\,
n(\xx)\,
\mathbf{1}^{H}_\alpha(\xx),
\nonumber\\
\bar n^T_\beta
&=&
\frac{1}{V}
\int_V d^3x\,
n(\xx)\,
\mathbf{1}^{T}_\beta(\xx),
\label{eq:environment_number_densities}
\end{eqnarray}
and the corresponding fractions
\begin{equation}
f^H_\alpha=\frac{\bar n^H_\alpha}{\bar n},
\qquad
f^T_\beta=\frac{\bar n^T_\beta}{\bar n}.
\end{equation}
The four numbers ${f_K,f_F,f_S,f_V}$ quantify how the tracer population
is distributed among knots, filaments, sheets, and voids.  They are used
to normalize the weighted catalog so that the mean tracer abundance is
preserved.  For the joint Hessian--tidal weighting this gives
\begin{equation}
\bar n_{wv}
=
\bar n
\sum_{\alpha,\beta}
f^{HT}_{\alpha\beta}\,
w_\alpha v_\beta,
\label{eq:joint_normalization}
\end{equation}
where $f^{HT}_{\alpha\beta}$ is the joint fraction of tracers classified
as $\alpha$ by $\mathcal{H}_{ij}$ and as $\beta$ by
$\mathcal{T}_{ij}$.  When the two operators are fitted separately, the
corresponding one-dimensional normalizations are
\begin{equation}
\sum_{\alpha} f^H_\alpha w_\alpha = 1,
\qquad
\sum_{\beta} f^T_\beta v_\beta = 1.
\label{eq:one_dimensional_normalizations}
\end{equation}
{These $w_\alpha$ are response weights for transforming a
\emph{fixed Gaussian reference catalogue}, rather than absolute tracer
occupation weights in the Gaussian cosmic web. Their common value at
$f_{\rm NL}=0$ is a convention of this transformation. The unweighted
number densities of Gaussian tracers in knots, filaments, sheets, and
voids need not be equal, and generally are not. An observational
estimator must therefore compare its measured environment-dependent
weights with the corresponding Gaussian baseline, not test equality of
raw environmental occupations.}

The weighted overdensities are defined by
\begin{equation}
\delta_w(\xx)
=
\frac{n_w(\xx)}{\bar n_w}-1,
\qquad
\delta_{wv}(\xx)
=
\frac{n_{wv}(\xx)}{\bar n_{wv}}-1.
\end{equation}
We determine $\bm{w}$, and when included $\bm{v}$, by fitting the
large-scale ratio
\begin{equation}
R(k)
=
\frac{P_{wm}(k)}{P_{hm}^{G}(k)}
\end{equation}
or, equivalently, by matching the non-Gaussian response
\begin{equation}
\Delta b(k)
=
\frac{P_{hm}^{\rm PNG}(k)}
{P_{mm}^{\rm PNG}(k)}
-
\frac{P_{hm}^{G}(k)}
{P_{mm}^{G}(k)}.
\end{equation}
The target model is
\begin{equation}
\Delta b(k)
=
A_{\rm PNG}
\mathcal{M}^{-1}(k,z),
\qquad
A_{\rm PNG}\equiv \fnl b_\Phi.
\label{eq:fitmodel}
\end{equation}

{The environment-dependent formulation further suggests a
direct observable for future surveys. Let $\bm{u}$ denote survey-calibrated
tracer occupation or relative-bias weights, with Gaussian reference
$\bm{u}^{(0)}$. Unlike the response factors $\bm w$ defined above,
$\bm{u}^{(0)}$ need not have equal components. At leading order, the
PNG-specific contrast obeys
\begin{equation}
\Delta\bm{u}_{\rm PNG}
=\bm{u}-\bm{u}^{(0)}
\simeq\bm{\mathcal C}f_{\rm NL},
\label{eq:environmental_response}
\end{equation}
where $\bm{\mathcal C}$ is calibrated from simulations with known
$f_{\rm NL}$. In the independent, normalized-contrast subspace, an
inverse-covariance estimator can be constructed as
\begin{equation}
\widehat f_{\rm NL}^{\rm web}=
\frac{\bm{\mathcal C}^{\rm T}\mathbf\Sigma_u^+
(\widehat{\bm u}-\bm u^{(0)})}
{\bm{\mathcal C}^{\rm T}\mathbf\Sigma_u^+
\bm{\mathcal C}},
\label{eq:environmental_estimator}
\end{equation}
where $\mathbf\Sigma_u^+$ denotes the inverse on the independent
contrast subspace. This construction is a proposed follow-up estimator,
not a measurement of its eventual $f_{\rm NL}$ precision. In
particular, it does not by itself remove the $b_\Phi$ degeneracy, and
no universal $\sqrt{2}$ sensitivity gain follows from the relation
$P_{hh}\propto b^2$.}

\paragraph{Simulation test.}
We test this construction with the \textsc{AbacusPNG} suite of local-PNG
$N$-body simulations \cite{2024PhRvD.109j3530H}; related local-PNG
simulation work has emphasized the mass dependence of halo clustering and
accurate treatment of initial conditions and aliasing
\cite{2024A&A...689A..69G,2026A&A...705A..20A}.  We use the simulations with
$f_{\rm NL}=\{-100,-30,30,100\}$, {analysing the
$z=1.1$ snapshot,} which allow us to test both the sign and
amplitude dependence of the web response.  For each catalog, we compute
the smoothed density field, construct $\mathcal{H}_{ij}$, classify the
grid into the four large-scale web regions, and assign the corresponding
weight to halos.  We then measure the auto- and cross-power spectra of the
weighted and unweighted halo catalogs.

\begin{figure}[tbp]
\centering
\includegraphics[width=\linewidth, height=1.2\textheight, keepaspectratio]{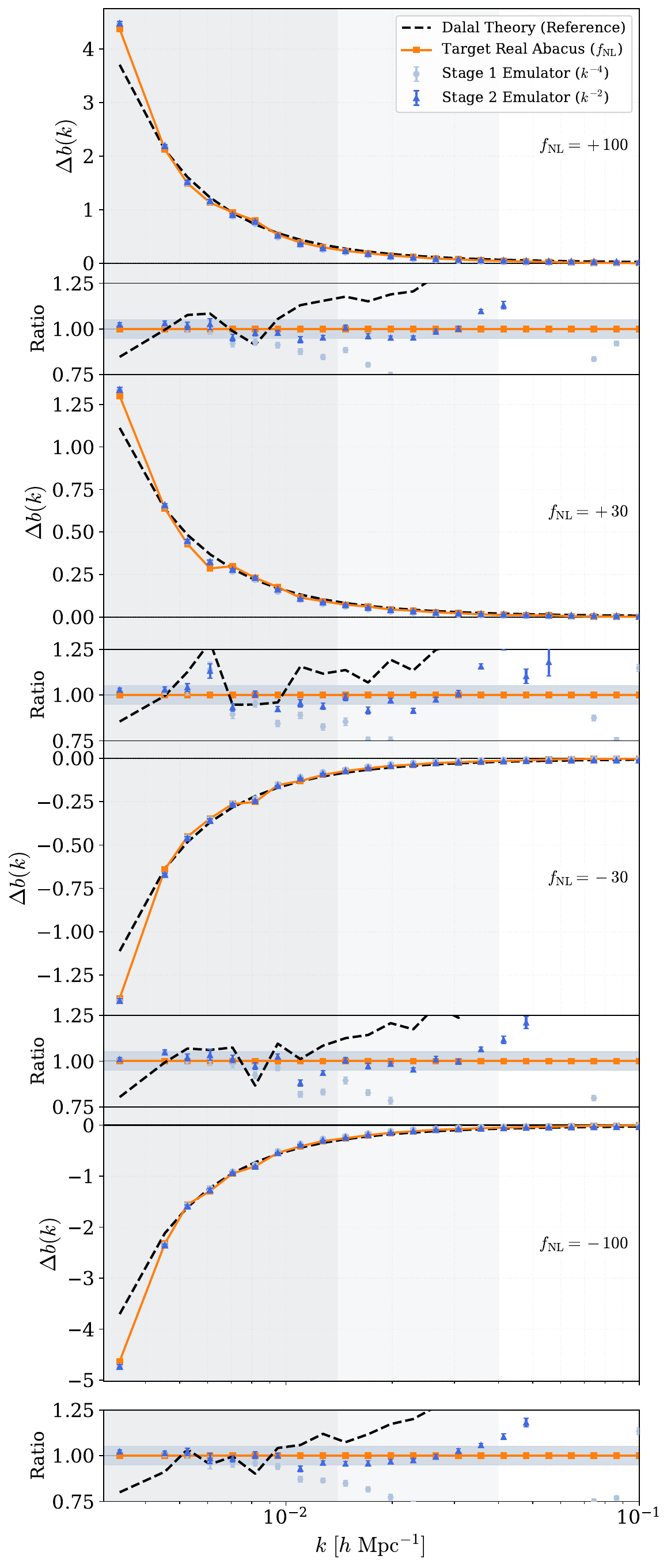}
\caption{{Large-scale bias emulation (blue circles) for different configurations vs reference calculations (orange squares) and Dalal et al. formula (dashed curves).}
   The dark
grey band marks the range dominated by the $-k^{-4}\delta$ Hessian
emulation, while the light grey band marks the range where the
$-k^{-2}\delta$ tidal correction improves the agreement.  The
horizontal bars in the ratio panels indicate $\pm 5\%$. Error bars are based on 20 \textsc{Abacus } simulations.}
\label{fig:power}
\end{figure}

\begin{figure}[h!] 
\centering
\includegraphics[width=\columnwidth, keepaspectratio]{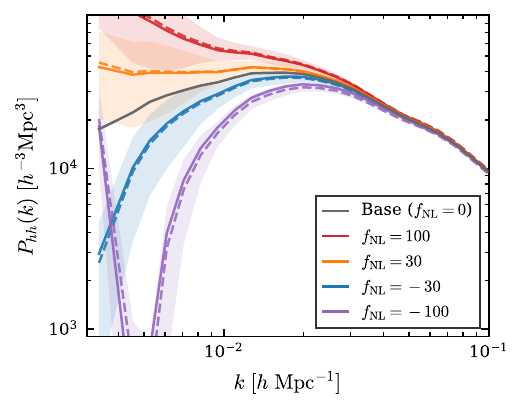}
\caption{Shown are the emulated (dashed lines) and the corresponding reference (solid lines) power-spectra for different $f_{\rm NL}$ values. Shaded areas correspond to 20 \textsc{Abacus } simulations.}
\label{fig:power_spectrum}
\end{figure}

Figure~\ref{fig:power} shows the central result.  The unweighted Gaussian
catalog has an approximately scale-independent large-scale bias, while the
local-PNG catalogs exhibit the expected rise or suppression toward small
$k$, depending on the sign of $f_{\rm NL}$.  The web-weighted halo
catalog reproduces this response without modifying the particle dynamics
or regenerating the initial conditions. {As illustrated by the comparison with the
Dalal et al.\ leading-order reference in Fig.~\ref{fig:power}, the calibrated
cosmic-web response tracks the $N$-body-measured PNG bias more closely
across the finite wavenumber range displayed. The method thereby models
more than the asymptotic $k^{-2}$ dependence: it captures the measured
scale dependence through a small set of environmental occupations.}  The agreement demonstrates that
the leading local-PNG signature can be represented as a nonlocal
large-scale environmental modulation. {The power-spectrum
comparison is the calibration and validation demonstrated here; independent
realizations, intermediate $f_{\rm NL}$ values, and PNG-specific squeezed
bispectra provide further discriminating tests. Their interest should not
be confused with an intrinsic inability of nonlocal bias models to
reproduce higher-order statistics.} Figure~\ref{fig:power_spectrum} shows that the emulated PNG power spectra are statistically consistent with the corresponding $N$-body measurements, with deviations remaining within the cosmic-variance uncertainty.
The weights can also be interpreted operationally as a cosmic-web-dependent
selection function.  Since only relative weights affect the environmental
modulation, we normalize them to the largest weight,
\begin{equation}
    \tilde w_\alpha
    =
    \frac{w_\alpha}{\max_{\beta} w_\beta},
    \qquad
    0\leq \tilde w_\alpha \leq 1 .
\end{equation}
A PNG-like catalog can then be generated by downsampling objects in each
web class with probability $\tilde w_\alpha$.  In this way, a single
high-resolution $N$-body simulation can be converted into a family of
effective PNG catalogs with different values and signs of $f_{\rm NL}$,
by changing only the large-scale cosmic-web-dependent selection
probabilities.

\begin{figure*}[tbp]
\centering
\includegraphics[width=\textwidth, height=0.55\textheight, keepaspectratio]{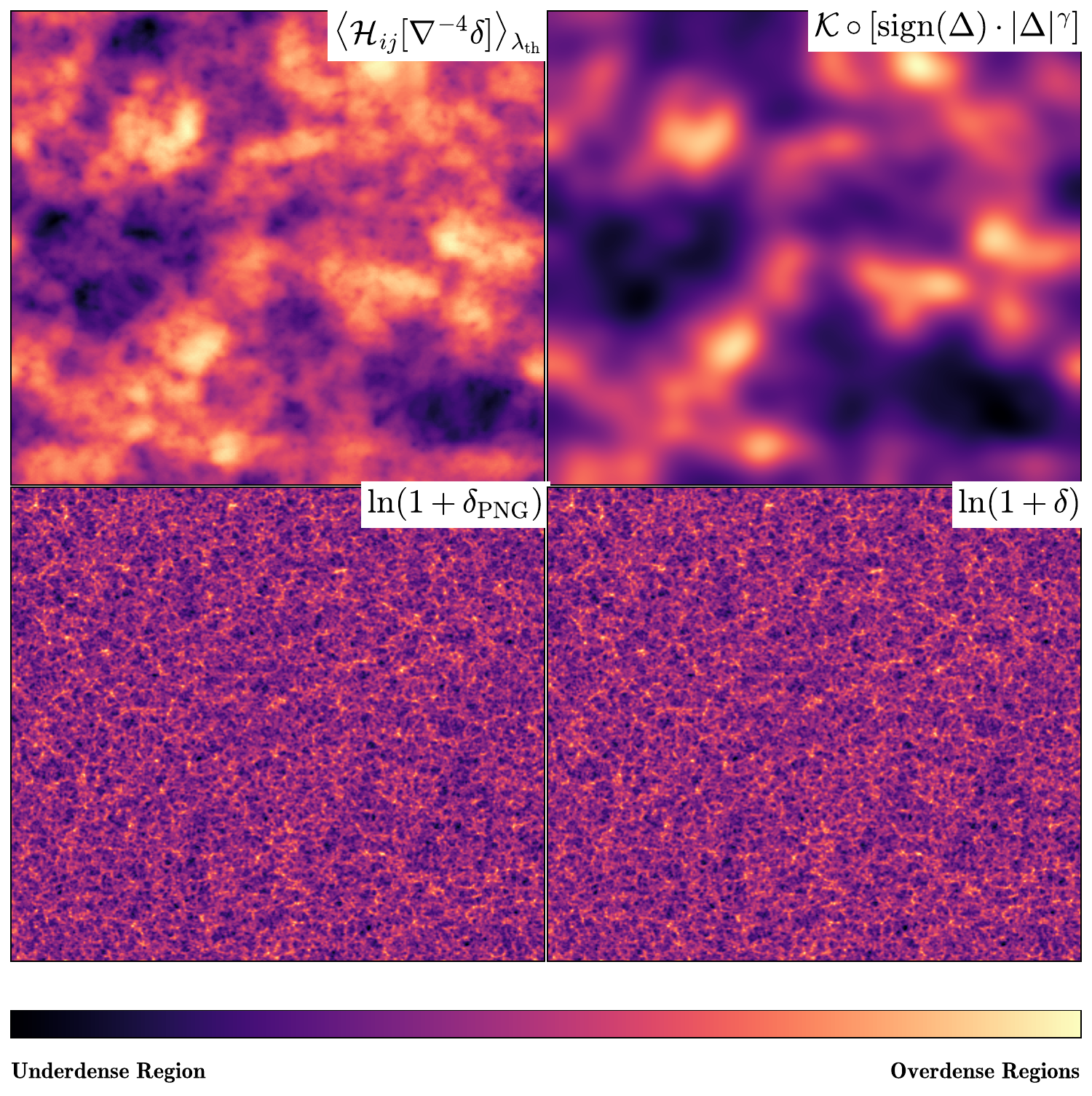}
\caption{
\textbf{Two-dimensional slices of the dark matter distribution and
its large-scale web classification.}
\textit{Bottom panels:} Two-dimensional sample slices of the dark
matter overdensity field, with side length
$2000\,h^{-1}\mathrm{Mpc}$ and grid size $N=360$, extracted from
\texttt{ABACUS} mock catalogs at $z=1.1$.  The left panel displays a
simulation with primordial non-Gaussianity implemented in the initial
conditions with $f_{\mathrm{NL}}=100$, while the right panel shows the
corresponding Gaussian catalog with the same seed.
\textit{Top panels:} Cosmic-web classifications obtained from the
$-k^{-4}\delta$ Hessian operator.  The structured appearance results from averaging over
different $\lambda_{\mathrm{th}}$ thresholds, ranging from $-250$ to
$250$.  The matter distribution is segmented into four large-scale
environments: knots, filaments, sheets, and voids.  The difference
field $\Delta\equiv \delta_{\rm PNG}-\delta$ is scaled to enhance the contrast, we also applied a gaussian filter $\mathcal{K}$
.}
\label{fig:weights}
\end{figure*}

\paragraph{Weight hierarchy.}
The fitted Hessian weights are not arbitrary.  We find a robust hierarchy
controlled by the sign of $\fnl$:
\begin{equation}
\fnl>0:
\qquad
w_K>w_F>w_S>w_V,
\label{eq:positivehierarchy}
\end{equation}
whereas
\begin{equation}
\fnl<0:
\qquad
 w_K<w_F<w_S<w_V.
\label{eq:negativehierarchy}
\end{equation}
This ordering has a simple interpretation.  Positive local PNG enhances
the abundance of biased tracers in regions selected by the corresponding
sign of the long-wavelength PNG modulation.  In the large-scale web
basis, this modulation is captured preferentially by knot-like
environments.  Negative $\fnl$ reverses the modulation, assigning the
largest relative weights to void-like environments.
This results is fully consistent with the well-known enhancements of knots in positive fnl and of voids in negative fnl scenarios \citep[][]{2008PhRvD..77l3514D}.
\begin{figure}[H] 
\centering
\includegraphics[width=\columnwidth, keepaspectratio]{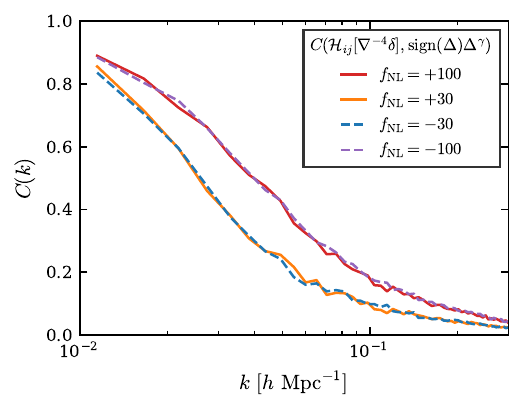}
\caption{We show the cross correlation of the averaged Hessian-Web over a wide range of eigenvalue thresholds and the difference field $\Delta = \delta_{\rm PNG}-\delta$ with $\gamma = 0.3$.}
\label{fig:CROSS}
\end{figure}
Table~\ref{tab:fnl_results} shows the fitted weights for positive and
negative $\fnl$.  The monotonic ordering of $\bm{w}$ is a compact,
interpretable summary of the PNG response.  The $\bm{v}$ rows quantify the
additional improvement obtained from the tidal-field tensor.  While the
Hessian weights $\bm{w}$ dominate the largest-scale $k^{-2}$ PNG response,
the tidal weights $\bm{v}$ absorb residual anisotropic environmental
dependence and improve the agreement at higher $k$.  The four
environment number-density fractions enter through the normalization of
the weighted catalog, ensuring that the reweighted sample preserves the
original mean abundance.
The cross-power spectrum between the $\mathcal{H}_{ij}$-based web field and the halo-density response induced by primordial non-Gaussianity reveals a clear large-scale correlation, as shown in Fig.~\ref{fig:CROSS}. 
The same correspondence is visually apparent in the field-level comparison shown in Fig.~\ref{fig:weights}.

\paragraph{Post-processing existing catalogs.}
An immediate application is to imprint an effective PNG signature on
existing Gaussian catalogs.  Given a catalog and its density field, one
computes $\mathcal{H}_{ij}$, assigns each object to a large-scale web
class, and applies the weights in Eq.~\eqref{eq:weightedfield_hessian}.
When extending the reconstruction to higher wavenumbers, one additionally
computes $\mathcal{T}_{ij}$ and applies the joint weighting in
Eq.~\eqref{eq:weightedfield_hessian_tidal}.  This operation changes the
large-scale clustering while preserving the small-scale catalog
structure, positions, velocities, and selection function.  It therefore
provides a fast emulator for PNG-like responses in mock catalogs.

{The essential advantage of applying these weights to a
high-resolution Gaussian $N$-body catalogue is that the detailed halo
phase-space distribution and the higher-order clustering already encoded
by the expensive reference simulation remain the starting point of the
construction. The PNG signal is imposed by adjusting the relative tracer
number densities within a few nonlocal web environments, rather than by
replacing the resolved catalogue with a new, substantially lower-resolution
evolution. These changes in occupation do modify the $N$-point statistics
of the resulting tracer sample; they do not erase the information on
halo positions, velocities and internal structure supplied by the
reference simulation. If the weights are implemented by stochastic
subsampling, the associated sampling noise must be included in any
covariance analysis.}


{The present operation does not regenerate
PNG-dependent merger histories or a fully evolved PNG halo mass function
\cite{2012PhRvD..85h3002S}; it is not intended to do so. This does
not imply a loss of the reference simulation's validated higher-order
clustering information. Indeed, the DESI mock challenge of
Balaguera-Antol\'{i}nez et al. \citep[][]{Balaguera_2023} 
validated halo abundance, two- and three-point statistics, and relevant
covariance structure against detailed $N$-body references.
Coloma-Nadal et al. \cite{2024JCAP...07..083C}
further demonstrated accurate bispectra from hierarchical cosmic-web
bias without fitting that observable directly. These tests underpin
our DESI \texttt{HOLI}-mocks  modelling programme and contradict any general
assertion that cosmic-web-based bias modelling is intrinsically
inaccurate at higher order. The newly introduced PNG weighting changes
the higher-order statistics of the output sample, as a PNG response
should; direct validation of its PNG-specific squeezed bispectrum and
covariance is complementary to the already demonstrated large-scale
halo--matter bias response.}


{
A promising direction for future work is to combine this
framework with machine-learning techniques to learn an
optimal mapping between high-fidelity $N$-body simulations
and computationally inexpensive PNG mock catalogues.
Such a multi-fidelity approach could learn the dependence
of nonlocal environmental bias on $f_{\rm NL}$, redshift,
tracer population, and cosmic-web morphology, while
retaining the detailed halo dynamics and higher-order
clustering information of the reference simulations.
This would extend the present low-dimensional weighting
scheme towards a flexible, physically interpretable
emulator, calibrated and validated against high-fidelity
simulations, for the efficient production of large
ensembles of PNG mocks.
}

\begin{table*}[t]
\caption{\label{tab:fnl_results}
Final results for the four Hessian-environment weights $\bm{w}$,
associated with the $-k^{-4}\delta$ PNG-like response, and the four
tidal-environment weights $\bm{v}$, associated with the $-k^{-2}\delta$
tidal-field correction.  The environment number-density fractions for
knots, filaments, sheets, and voids are used to normalize the weighted
catalog so that the mean tracer abundance is preserved.  For all
configurations, the average Gaussian bias is $b_1=2.2075$, measured from
the \texttt{ABACUS} simulations.
}
\begin{ruledtabular}
\begin{tabular}{cccccc}
$f_{\mathrm{NL}}$ & Config. & Knot & Filament & Sheet & Void \\
\hline
$100$  & $\bm{w}$ & $1.0000$ & $0.9779 \pm 0.0030$ & $0.9555 \pm 0.0031$ & $0.9283 \pm 0.0038$ \\
& $\bm{v}$ & $0.8217 \pm 0.1377$ & $0.9835 \pm 0.0986$ & $1.0000$ & $0.9489 \pm 0.0751$ \\
\hline
$30$   & $\bm{w}$ & $1.0000$ & $0.9945 \pm 0.0011$ & $0.9867 \pm 0.0011$ & $0.9790 \pm 0.0022$ \\
& $\bm{v}$ & $0.8867 \pm 0.0533$ & $0.9695 \pm 0.0283$ & $1.0000$ & $0.9677 \pm 0.0185$ \\
\hline
$-30$  & $\bm{w}$ & $0.9758 \pm 0.0021$ & $0.9835 \pm 0.0022$ & $0.9898 \pm 0.0021$ & $1.0000$ \\
& $\bm{v}$ & $0.9734 \pm 0.0502$ & $0.9814 \pm 0.0254$ & $1.0000$ & $0.9978 \pm 0.0200$ \\
\hline
$-100$ & $\bm{w}$ & $0.9230 \pm 0.0036$ & $0.9494 \pm 0.0033$ & $0.9698 \pm 0.0040$ & $1.0000$ \\
& $\bm{v}$ & $0.9156 \pm 0.1208$ & $0.9741 \pm 0.0614$ & $0.9853 \pm 0.0510$ & $1.0000$ \\
\end{tabular}
\end{ruledtabular}
\end{table*}

\paragraph{Discussion.}
We have shown that the leading large-scale signature of local primordial
non-Gaussianity can be expressed as a nonlocal cosmic-web bias.  The trace
of the Hessian $\partial_i\partial_j\nabla^{-4}\delta$ is
$\nabla^{-2}\delta$, and therefore has the same large-scale $k^{-2}$
behavior as the Dalal et al. scale-dependent-bias correction.
Diagonalizing this Hessian yields four large-scale environments.
Weighting tracers by these environments reproduces the PNG-induced rise
of the bias and reveals a sign-dependent hierarchy among knot-,
filament-, sheet-, and void-like regions.

The inclusion of the tidal tensor
$\mathcal{T}_{ij}=\partial_i\partial_j\nabla^{-2}\delta$ extends the
description beyond the leading PNG-like response.  In contrast to
$\mathcal{H}_{ij}$, whose trace carries the inverse-Laplacian enhancement,
$\mathcal{T}_{ij}$ describes the standard anisotropic tidal environment.
Its four additional weights capture residual environmental dependence and
improve the reconstruction toward higher wavenumbers while preserving a
compact eight-parameter structure.

This perspective connects three ideas that are usually treated
separately: primordial mode coupling, scale-dependent bias, and
anisotropic environmental dependence.  It also provides a fast way to
post-process halo or galaxy mock catalogs into catalogs with an effective
large-scale PNG response, while preserving their small-scale structure,
positions, velocities, and selection function.

{The new interpretation also motivates a complementary
observational programme. A standard PNG power-spectrum analysis is
particularly exposed to contamination of its lowest Fourier modes. In
contrast, a cosmic-web measurement can compress the information into
tracer-number-density or relative-bias contrasts across the many knot-,
filament-, sheet-, and void-like regions distributed throughout the survey.
The observable is not the power at one chosen small $k$ but a calibrated
change in environmental occupation relative to Gaussian expectations.
Consequently, selection effects can be diagnosed and marginalized using
environment-conditioned counts, angular and radial splits, and survey
randoms propagated through the same web-reconstruction pipeline. The
inverse-Laplacian operator itself is sensitive to large-scale
contamination, so these are distinct controls rather than a guarantee
of immunity to systematics. A particularly promising strategy is a joint
analysis of the conventional $P(k)$ and the environmental contrasts,
with their cross-covariance modelled consistently.}

The observational use of this large-scale web response for PNG detection requires a
dedicated treatment of survey masks, selection functions, redshift-space
distortions, random catalogs, and calibration against mocks.  We leave
this data-analysis framework to a companion paper.

\begin{acknowledgments}
KPV thanks the IAC for hospitality during his undergraduate internship and bachelor thesis 
under the supervision of FSK and FS.  FSK proposed the idea of this project and acknowledges the Instituto de
Astrofísica de Canarias (IAC) for continuous support of the
\textit{Cosmology with LSS probes} activities.  The authors thank the
Spanish Ministry of Innovation and Science / Agencia Estatal de
Investigación for support of the project \textit{Big Data of the Cosmic
Web} (PID2020-120612GB-I00/AEI/10.13039/501100011033), under which this
work was conceived and carried out, and the \textit{FIRE (Field-level
bayesian Inference to Reconstruct the univErse)} project, recently funded
by a \textit{Proyecto de Generación de Conocimiento 2024},
PID2024-160504NB-I00.  FS acknowledges support from the \textit{Institute
for Fundamental Physics of the Universe} postdoctoral fellowship scheme.
FS is also grateful to IAC for hospitality during part of the realization
of this project.
\end{acknowledgments}

\bibliographystyle{apsrev4-2}
\bibliography{lit}

\begin{thebibliography}{23}%
\makeatletter
\providecommand \@ifxundefined [1]{%
 \@ifx{#1\undefined}
}%
\providecommand \@ifnum [1]{%
 \ifnum #1\expandafter \@firstoftwo
 \else \expandafter \@secondoftwo
 \fi
}%
\providecommand \@ifx [1]{%
 \ifx #1\expandafter \@firstoftwo
 \else \expandafter \@secondoftwo
 \fi
}%
\providecommand \natexlab [1]{#1}%
\providecommand \enquote  [1]{``#1''}%
\providecommand \bibnamefont  [1]{#1}%
\providecommand \bibfnamefont [1]{#1}%
\providecommand \citenamefont [1]{#1}%
\providecommand \href@noop [0]{\@secondoftwo}%
\providecommand \href [0]{\begingroup \@sanitize@url \@href}%
\providecommand \@href[1]{\@@startlink{#1}\@@href}%
\providecommand \@@href[1]{\endgroup#1\@@endlink}%
\providecommand \@sanitize@url [0]{\catcode `\\12\catcode `\$12\catcode `\&12\catcode `\#12\catcode `\^12\catcode `\_12\catcode `\%12\relax}%
\providecommand \@@startlink[1]{}%
\providecommand \@@endlink[0]{}%
\providecommand \url  [0]{\begingroup\@sanitize@url \@url }%
\providecommand \@url [1]{\endgroup\@href {#1}{\urlprefix }}%
\providecommand \urlprefix  [0]{URL }%
\providecommand \Eprint [0]{\href }%
\providecommand \doibase [0]{https://doi.org/}%
\providecommand \selectlanguage [0]{\@gobble}%
\providecommand \bibinfo  [0]{\@secondoftwo}%
\providecommand \bibfield  [0]{\@secondoftwo}%
\providecommand \translation [1]{[#1]}%
\providecommand \BibitemOpen [0]{}%
\providecommand \bibitemStop [0]{}%
\providecommand \bibitemNoStop [0]{.\EOS\space}%
\providecommand \EOS [0]{\spacefactor3000\relax}%
\providecommand \BibitemShut  [1]{\csname bibitem#1\endcsname}%
\let\auto@bib@innerbib\@empty
\bibitem [{\citenamefont {{Bartolo}}\ \emph {et~al.}(2004)\citenamefont {{Bartolo}}, \citenamefont {{Komatsu}}, \citenamefont {{Matarrese}},\ and\ \citenamefont {{Riotto}}}]{2004PhR...402..103B}%
  \BibitemOpen
  \bibfield  {author} {\bibinfo {author} {\bibfnamefont {N.}~\bibnamefont {{Bartolo}}}, \bibinfo {author} {\bibfnamefont {E.}~\bibnamefont {{Komatsu}}}, \bibinfo {author} {\bibfnamefont {S.}~\bibnamefont {{Matarrese}}},\ and\ \bibinfo {author} {\bibfnamefont {A.}~\bibnamefont {{Riotto}}},\ }\href {https://doi.org/10.1016/j.physrep.2004.08.022} {\bibfield  {journal} {\bibinfo  {journal} {Physics Reports}\ }\textbf {\bibinfo {volume} {402}},\ \bibinfo {pages} {103} (\bibinfo {year} {2004})},\ \Eprint {https://arxiv.org/abs/astro-ph/0406398} {arXiv:astro-ph/0406398 [astro-ph]} \BibitemShut {NoStop}%
\bibitem [{\citenamefont {{Chen}}(2010)}]{2010AdAst2010E..72C}%
  \BibitemOpen
  \bibfield  {author} {\bibinfo {author} {\bibfnamefont {X.}~\bibnamefont {{Chen}}},\ }\href {https://doi.org/10.1155/2010/638979} {\bibfield  {journal} {\bibinfo  {journal} {Advances in Astronomy}\ }\textbf {\bibinfo {volume} {2010}},\ \bibinfo {eid} {638979} (\bibinfo {year} {2010})},\ \Eprint {https://arxiv.org/abs/1002.1416} {arXiv:1002.1416 [astro-ph.CO]} \BibitemShut {NoStop}%
\bibitem [{\citenamefont {{Maldacena}}(2003)}]{2003JHEP...05..013M}%
  \BibitemOpen
  \bibfield  {author} {\bibinfo {author} {\bibfnamefont {J.}~\bibnamefont {{Maldacena}}},\ }\href {https://doi.org/10.1088/1126-6708/2003/05/013} {\bibfield  {journal} {\bibinfo  {journal} {Journal of High Energy Physics}\ }\textbf {\bibinfo {volume} {2003}},\ \bibinfo {eid} {013} (\bibinfo {year} {2003})},\ \Eprint {https://arxiv.org/abs/astro-ph/0210603} {arXiv:astro-ph/0210603 [astro-ph]} \BibitemShut {NoStop}%
\bibitem [{\citenamefont {{Acquaviva}}\ \emph {et~al.}(2003)\citenamefont {{Acquaviva}}, \citenamefont {{Bartolo}}, \citenamefont {{Matarrese}},\ and\ \citenamefont {{Riotto}}}]{2003NuPhB.667..119A}%
  \BibitemOpen
  \bibfield  {author} {\bibinfo {author} {\bibfnamefont {V.}~\bibnamefont {{Acquaviva}}}, \bibinfo {author} {\bibfnamefont {N.}~\bibnamefont {{Bartolo}}}, \bibinfo {author} {\bibfnamefont {S.}~\bibnamefont {{Matarrese}}},\ and\ \bibinfo {author} {\bibfnamefont {A.}~\bibnamefont {{Riotto}}},\ }\href {https://doi.org/10.1016/S0550-3213(03)00550-9} {\bibfield  {journal} {\bibinfo  {journal} {Nuclear Physics B}\ }\textbf {\bibinfo {volume} {667}},\ \bibinfo {pages} {119} (\bibinfo {year} {2003})},\ \Eprint {https://arxiv.org/abs/astro-ph/0209156} {arXiv:astro-ph/0209156 [astro-ph]} \BibitemShut {NoStop}%
\bibitem [{\citenamefont {{Creminelli}}\ and\ \citenamefont {{Zaldarriaga}}(2004)}]{2004JCAP...10..006C}%
  \BibitemOpen
  \bibfield  {author} {\bibinfo {author} {\bibfnamefont {P.}~\bibnamefont {{Creminelli}}}\ and\ \bibinfo {author} {\bibfnamefont {M.}~\bibnamefont {{Zaldarriaga}}},\ }\href {https://doi.org/10.1088/1475-7516/2004/10/006} {\bibfield  {journal} {\bibinfo  {journal} {Journal of Cosmology and Astroparticle Physics}\ }\textbf {\bibinfo {volume} {2004}}\bibfield  {number} {\bibinfo  {number} { (10)},\ \bibinfo {eid} {006}},\ }\Eprint {https://arxiv.org/abs/astro-ph/0407059} {arXiv:astro-ph/0407059 [astro-ph]} \BibitemShut {NoStop}%
\bibitem [{\citenamefont {{Planck Collaboration}}(2020)}]{2020A&A...641A...9P}%
  \BibitemOpen
  \bibfield  {author} {\bibinfo {author} {\bibnamefont {{Planck Collaboration}}},\ }\href {https://doi.org/10.1051/0004-6361/201935891} {\bibfield  {journal} {\bibinfo  {journal} {Astronomy \& Astrophysics}\ }\textbf {\bibinfo {volume} {641}},\ \bibinfo {eid} {A9} (\bibinfo {year} {2020})},\ \Eprint {https://arxiv.org/abs/1905.05697} {arXiv:1905.05697 [astro-ph.CO]} \BibitemShut {NoStop}%
\bibitem [{\citenamefont {{Byrnes}}\ and\ \citenamefont {{Choi}}(2010)}]{2010AdAst2010E..76B}%
  \BibitemOpen
  \bibfield  {author} {\bibinfo {author} {\bibfnamefont {C.~T.}\ \bibnamefont {{Byrnes}}}\ and\ \bibinfo {author} {\bibfnamefont {K.-Y.}\ \bibnamefont {{Choi}}},\ }\href {https://doi.org/10.1155/2010/724525} {\bibfield  {journal} {\bibinfo  {journal} {Advances in Astronomy}\ }\textbf {\bibinfo {volume} {2010}},\ \bibinfo {eid} {724525} (\bibinfo {year} {2010})},\ \Eprint {https://arxiv.org/abs/1002.3110} {arXiv:1002.3110 [astro-ph.CO]} \BibitemShut {NoStop}%
\bibitem [{\citenamefont {{Dalal}}\ \emph {et~al.}(2008)\citenamefont {{Dalal}}, \citenamefont {{Dor{\'e}}}, \citenamefont {{Huterer}},\ and\ \citenamefont {{Shirokov}}}]{2008PhRvD..77l3514D}%
  \BibitemOpen
  \bibfield  {author} {\bibinfo {author} {\bibfnamefont {N.}~\bibnamefont {{Dalal}}}, \bibinfo {author} {\bibfnamefont {O.}~\bibnamefont {{Dor{\'e}}}}, \bibinfo {author} {\bibfnamefont {D.}~\bibnamefont {{Huterer}}},\ and\ \bibinfo {author} {\bibfnamefont {A.}~\bibnamefont {{Shirokov}}},\ }\href {https://doi.org/10.1103/PhysRevD.77.123514} {\bibfield  {journal} {\bibinfo  {journal} {Physical Review D}\ }\textbf {\bibinfo {volume} {77}},\ \bibinfo {eid} {123514} (\bibinfo {year} {2008})},\ \Eprint {https://arxiv.org/abs/0710.4560} {arXiv:0710.4560 [astro-ph]} \BibitemShut {NoStop}%
\bibitem [{\citenamefont {{Slosar}}\ \emph {et~al.}(2008)\citenamefont {{Slosar}}, \citenamefont {{Hirata}}, \citenamefont {{Seljak}}, \citenamefont {{Ho}},\ and\ \citenamefont {{Padmanabhan}}}]{2008JCAP...08..031S}%
  \BibitemOpen
  \bibfield  {author} {\bibinfo {author} {\bibfnamefont {A.}~\bibnamefont {{Slosar}}}, \bibinfo {author} {\bibfnamefont {C.}~\bibnamefont {{Hirata}}}, \bibinfo {author} {\bibfnamefont {U.}~\bibnamefont {{Seljak}}}, \bibinfo {author} {\bibfnamefont {S.}~\bibnamefont {{Ho}}},\ and\ \bibinfo {author} {\bibfnamefont {N.}~\bibnamefont {{Padmanabhan}}},\ }\href {https://doi.org/10.1088/1475-7516/2008/08/031} {\bibfield  {journal} {\bibinfo  {journal} {Journal of Cosmology and Astroparticle Physics}\ }\textbf {\bibinfo {volume} {2008}}\bibfield  {number} {\bibinfo  {number} { (8)},\ \bibinfo {eid} {031}},\ }\Eprint {https://arxiv.org/abs/0805.3580} {arXiv:0805.3580 [astro-ph]} \BibitemShut {NoStop}%
\bibitem [{\citenamefont {{Matarrese}}\ and\ \citenamefont {{Verde}}(2008)}]{2008ApJ...677L..77M}%
  \BibitemOpen
  \bibfield  {author} {\bibinfo {author} {\bibfnamefont {S.}~\bibnamefont {{Matarrese}}}\ and\ \bibinfo {author} {\bibfnamefont {L.}~\bibnamefont {{Verde}}},\ }\href {https://doi.org/10.1086/587840} {\bibfield  {journal} {\bibinfo  {journal} {Astrophysical Journal Letters}\ }\textbf {\bibinfo {volume} {677}},\ \bibinfo {pages} {L77} (\bibinfo {year} {2008})},\ \Eprint {https://arxiv.org/abs/0801.4826} {arXiv:0801.4826 [astro-ph]} \BibitemShut {NoStop}%
\bibitem [{\citenamefont {{Desjacques}}\ and\ \citenamefont {{Seljak}}(2010)}]{2010PhRvD..81b3006D}%
  \BibitemOpen
  \bibfield  {author} {\bibinfo {author} {\bibfnamefont {V.}~\bibnamefont {{Desjacques}}}\ and\ \bibinfo {author} {\bibfnamefont {U.}~\bibnamefont {{Seljak}}},\ }\href {https://doi.org/10.1103/PhysRevD.81.023006} {\bibfield  {journal} {\bibinfo  {journal} {Physical Review D}\ }\textbf {\bibinfo {volume} {81}},\ \bibinfo {eid} {023006} (\bibinfo {year} {2010})},\ \Eprint {https://arxiv.org/abs/0907.2257} {arXiv:0907.2257 [astro-ph.CO]} \BibitemShut {NoStop}%
\bibitem [{\citenamefont {{Hamaus}}\ \emph {et~al.}(2011)\citenamefont {{Hamaus}}, \citenamefont {{Seljak}},\ and\ \citenamefont {{Desjacques}}}]{2011PhRvD..84h3509H}%
  \BibitemOpen
  \bibfield  {author} {\bibinfo {author} {\bibfnamefont {N.}~\bibnamefont {{Hamaus}}}, \bibinfo {author} {\bibfnamefont {U.}~\bibnamefont {{Seljak}}},\ and\ \bibinfo {author} {\bibfnamefont {V.}~\bibnamefont {{Desjacques}}},\ }\href {https://doi.org/10.1103/PhysRevD.84.083509} {\bibfield  {journal} {\bibinfo  {journal} {Physical Review D}\ }\textbf {\bibinfo {volume} {84}},\ \bibinfo {eid} {083509} (\bibinfo {year} {2011})},\ \Eprint {https://arxiv.org/abs/1104.2321} {arXiv:1104.2321 [astro-ph.CO]} \BibitemShut {NoStop}%
\bibitem [{\citenamefont {{Karagiannis}}\ \emph {et~al.}(2018)\citenamefont {{Karagiannis}}, \citenamefont {{Lazanu}}, \citenamefont {{Liguori}}, \citenamefont {{Raccanelli}}, \citenamefont {{Bartolo}},\ and\ \citenamefont {{Verde}}}]{2018MNRAS.478.1341K}%
  \BibitemOpen
  \bibfield  {author} {\bibinfo {author} {\bibfnamefont {D.}~\bibnamefont {{Karagiannis}}}, \bibinfo {author} {\bibfnamefont {A.}~\bibnamefont {{Lazanu}}}, \bibinfo {author} {\bibfnamefont {M.}~\bibnamefont {{Liguori}}}, \bibinfo {author} {\bibfnamefont {A.}~\bibnamefont {{Raccanelli}}}, \bibinfo {author} {\bibfnamefont {N.}~\bibnamefont {{Bartolo}}},\ and\ \bibinfo {author} {\bibfnamefont {L.}~\bibnamefont {{Verde}}},\ }\href {https://doi.org/10.1093/mnras/sty1029} {\bibfield  {journal} {\bibinfo  {journal} {Monthly Notices of the Royal Astronomical Society}\ }\textbf {\bibinfo {volume} {478}},\ \bibinfo {pages} {1341} (\bibinfo {year} {2018})},\ \Eprint {https://arxiv.org/abs/1801.09280} {arXiv:1801.09280 [astro-ph.CO]} \BibitemShut {NoStop}%
\bibitem [{\citenamefont {{Ballardini}}\ \emph {et~al.}(2019)\citenamefont {{Ballardini}}, \citenamefont {{Matthewson}},\ and\ \citenamefont {{Maartens}}}]{2019MNRAS.489.1950B}%
  \BibitemOpen
  \bibfield  {author} {\bibinfo {author} {\bibfnamefont {M.}~\bibnamefont {{Ballardini}}}, \bibinfo {author} {\bibfnamefont {W.~L.}\ \bibnamefont {{Matthewson}}},\ and\ \bibinfo {author} {\bibfnamefont {R.}~\bibnamefont {{Maartens}}},\ }\href {https://doi.org/10.1093/mnras/stz2258} {\bibfield  {journal} {\bibinfo  {journal} {Monthly Notices of the Royal Astronomical Society}\ }\textbf {\bibinfo {volume} {489}},\ \bibinfo {pages} {1950} (\bibinfo {year} {2019})},\ \Eprint {https://arxiv.org/abs/1906.04730} {arXiv:1906.04730 [astro-ph.CO]} \BibitemShut {NoStop}%
\bibitem [{\citenamefont {{Riquelme}}\ \emph {et~al.}(2023)\citenamefont {{Riquelme}}, \citenamefont {{Avila}}, \citenamefont {{Garc{\'\i}a-Bellido}}, \citenamefont {{Porredon}}, \citenamefont {{Ferrero}}, \citenamefont {{Chan}}, \citenamefont {{Rosenfeld}}, \citenamefont {{Camacho}}, \citenamefont {{Adame}}, \citenamefont {{Carnero Rosell}}, \citenamefont {{Crocce}}, \citenamefont {{De Vicente}}, \citenamefont {{Eifler}}, \citenamefont {{Elvin-Poole}}, \citenamefont {{Fang}}, \citenamefont {{Krause}}, \citenamefont {{Rodriguez Monroy}}, \citenamefont {{Ross}}, \citenamefont {{Sanchez}},\ and\ \citenamefont {{Sevilla}}}]{2023MNRAS.523..603R}%
  \BibitemOpen
  \bibfield  {author} {\bibinfo {author} {\bibfnamefont {W.}~\bibnamefont {{Riquelme}}}, \bibinfo {author} {\bibfnamefont {S.}~\bibnamefont {{Avila}}}, \bibinfo {author} {\bibfnamefont {J.}~\bibnamefont {{Garc{\'\i}a-Bellido}}}, \bibinfo {author} {\bibfnamefont {A.}~\bibnamefont {{Porredon}}}, \bibinfo {author} {\bibfnamefont {I.}~\bibnamefont {{Ferrero}}}, \bibinfo {author} {\bibfnamefont {K.~C.}\ \bibnamefont {{Chan}}}, \bibinfo {author} {\bibfnamefont {R.}~\bibnamefont {{Rosenfeld}}}, \bibinfo {author} {\bibfnamefont {H.}~\bibnamefont {{Camacho}}}, \bibinfo {author} {\bibfnamefont {A.~G.}\ \bibnamefont {{Adame}}}, \bibinfo {author} {\bibfnamefont {A.}~\bibnamefont {{Carnero Rosell}}}, \bibinfo {author} {\bibfnamefont {M.}~\bibnamefont {{Crocce}}}, \bibinfo {author} {\bibfnamefont {J.}~\bibnamefont {{De Vicente}}}, \bibinfo {author} {\bibfnamefont {T.}~\bibnamefont {{Eifler}}}, \bibinfo {author} {\bibfnamefont {J.}~\bibnamefont {{Elvin-Poole}}}, \bibinfo {author} {\bibfnamefont {X.}~\bibnamefont {{Fang}}},
  \bibinfo {author} {\bibfnamefont {E.}~\bibnamefont {{Krause}}}, \bibinfo {author} {\bibfnamefont {M.}~\bibnamefont {{Rodriguez Monroy}}}, \bibinfo {author} {\bibfnamefont {A.~J.}\ \bibnamefont {{Ross}}}, \bibinfo {author} {\bibfnamefont {E.}~\bibnamefont {{Sanchez}}},\ and\ \bibinfo {author} {\bibfnamefont {I.}~\bibnamefont {{Sevilla}}},\ }\href {https://doi.org/10.1093/mnras/stad1429} {\bibfield  {journal} {\bibinfo  {journal} {Monthly Notices of the Royal Astronomical Society}\ }\textbf {\bibinfo {volume} {523}},\ \bibinfo {pages} {603} (\bibinfo {year} {2023})},\ \Eprint {https://arxiv.org/abs/2209.07187} {arXiv:2209.07187 [astro-ph.CO]} \BibitemShut {NoStop}%
\bibitem [{\citenamefont {{Heinrich}}\ \emph {et~al.}(2024)\citenamefont {{Heinrich}}, \citenamefont {{Dor{\'e}}},\ and\ \citenamefont {{Krause}}}]{2024PhRvD.109l3511H}%
  \BibitemOpen
  \bibfield  {author} {\bibinfo {author} {\bibfnamefont {C.}~\bibnamefont {{Heinrich}}}, \bibinfo {author} {\bibfnamefont {O.}~\bibnamefont {{Dor{\'e}}}},\ and\ \bibinfo {author} {\bibfnamefont {E.}~\bibnamefont {{Krause}}},\ }\href {https://doi.org/10.1103/PhysRevD.109.123511} {\bibfield  {journal} {\bibinfo  {journal} {Physical Review D}\ }\textbf {\bibinfo {volume} {109}},\ \bibinfo {eid} {123511} (\bibinfo {year} {2024})},\ \Eprint {https://arxiv.org/abs/2311.13082} {arXiv:2311.13082 [astro-ph.CO]} \BibitemShut {NoStop}%
\bibitem [{\citenamefont {{Kitaura}}\ and\ \citenamefont {{Sinigaglia}}(2026)}]{2026arXiv260315834K}%
  \BibitemOpen
  \bibfield  {author} {\bibinfo {author} {\bibfnamefont {F.-S.}\ \bibnamefont {{Kitaura}}}\ and\ \bibinfo {author} {\bibfnamefont {F.}~\bibnamefont {{Sinigaglia}}},\ }\href {https://doi.org/10.48550/arXiv.2603.15834} {\bibfield  {journal} {\bibinfo  {journal} {arXiv e-prints}\ ,\ \bibinfo {eid} {arXiv:2603.15834}} (\bibinfo {year} {2026})},\ \Eprint {https://arxiv.org/abs/2603.15834} {arXiv:2603.15834 [astro-ph.CO]} \BibitemShut {NoStop}%
\bibitem [{\citenamefont {{Coloma-Nadal}}\ \emph {et~al.}(2024)\citenamefont {{Coloma-Nadal}}, \citenamefont {{Kitaura}}, \citenamefont {{Garc{\'\i}a-Farieta}}, \citenamefont {{Sinigaglia}}, \citenamefont {{Favole}},\ and\ \citenamefont {{Forero S{\'a}nchez}}}]{2024JCAP...07..083C}%
  \BibitemOpen
  \bibfield  {author} {\bibinfo {author} {\bibfnamefont {J.~M.}\ \bibnamefont {{Coloma-Nadal}}}, \bibinfo {author} {\bibfnamefont {F.-S.}\ \bibnamefont {{Kitaura}}}, \bibinfo {author} {\bibfnamefont {J.~E.}\ \bibnamefont {{Garc{\'\i}a-Farieta}}}, \bibinfo {author} {\bibfnamefont {F.}~\bibnamefont {{Sinigaglia}}}, \bibinfo {author} {\bibfnamefont {G.}~\bibnamefont {{Favole}}},\ and\ \bibinfo {author} {\bibfnamefont {D.}~\bibnamefont {{Forero S{\'a}nchez}}},\ }\href {https://doi.org/10.1088/1475-7516/2024/07/083} {\bibfield  {journal} {\bibinfo  {journal} {Journal of Cosmology and Astroparticle Physics}\ }\textbf {\bibinfo {volume} {2024}}\bibfield  {number} {\bibinfo  {number} { (7)},\ \bibinfo {eid} {083}},\ }\Eprint {https://arxiv.org/abs/2403.19337} {arXiv:2403.19337 [astro-ph.CO]} \BibitemShut {NoStop}%
\bibitem [{\citenamefont {{Hadzhiyska}}\ \emph {et~al.}(2024)\citenamefont {{Hadzhiyska}}, \citenamefont {{Garrison}}, \citenamefont {{Eisenstein}},\ and\ \citenamefont {{Ferraro}}}]{2024PhRvD.109j3530H}%
  \BibitemOpen
  \bibfield  {author} {\bibinfo {author} {\bibfnamefont {B.}~\bibnamefont {{Hadzhiyska}}}, \bibinfo {author} {\bibfnamefont {L.~H.}\ \bibnamefont {{Garrison}}}, \bibinfo {author} {\bibfnamefont {D.~J.}\ \bibnamefont {{Eisenstein}}},\ and\ \bibinfo {author} {\bibfnamefont {S.}~\bibnamefont {{Ferraro}}},\ }\href {https://doi.org/10.1103/PhysRevD.109.103530} {\bibfield  {journal} {\bibinfo  {journal} {Physical Review D}\ }\textbf {\bibinfo {volume} {109}},\ \bibinfo {eid} {103530} (\bibinfo {year} {2024})},\ \Eprint {https://arxiv.org/abs/2402.10881} {arXiv:2402.10881 [astro-ph.CO]} \BibitemShut {NoStop}%
\bibitem [{\citenamefont {{Guti{\'e}rrez Adame}}\ \emph {et~al.}(2024)\citenamefont {{Guti{\'e}rrez Adame}}, \citenamefont {{Avila}}, \citenamefont {{Gonzalez-Perez}}, \citenamefont {{Yepes}}, \citenamefont {{Pellejero}}, \citenamefont {{Wang}}, \citenamefont {{Chuang}}, \citenamefont {{Feng}}, \citenamefont {{Garcia-Bellido}},\ and\ \citenamefont {{Knebe}}}]{2024A&A...689A..69G}%
  \BibitemOpen
  \bibfield  {author} {\bibinfo {author} {\bibfnamefont {A.}~\bibnamefont {{Guti{\'e}rrez Adame}}}, \bibinfo {author} {\bibfnamefont {S.}~\bibnamefont {{Avila}}}, \bibinfo {author} {\bibfnamefont {V.}~\bibnamefont {{Gonzalez-Perez}}}, \bibinfo {author} {\bibfnamefont {G.}~\bibnamefont {{Yepes}}}, \bibinfo {author} {\bibfnamefont {M.}~\bibnamefont {{Pellejero}}}, \bibinfo {author} {\bibfnamefont {M.~S.}\ \bibnamefont {{Wang}}}, \bibinfo {author} {\bibfnamefont {C.-H.}\ \bibnamefont {{Chuang}}}, \bibinfo {author} {\bibfnamefont {Y.}~\bibnamefont {{Feng}}}, \bibinfo {author} {\bibfnamefont {J.}~\bibnamefont {{Garcia-Bellido}}},\ and\ \bibinfo {author} {\bibfnamefont {A.}~\bibnamefont {{Knebe}}},\ }\href {https://doi.org/10.1051/0004-6361/202349037} {\bibfield  {journal} {\bibinfo  {journal} {Astronomy \& Astrophysics}\ }\textbf {\bibinfo {volume} {689}},\ \bibinfo {eid} {A69} (\bibinfo {year} {2024})},\ \Eprint {https://arxiv.org/abs/2312.12405} {arXiv:2312.12405 [astro-ph.CO]} \BibitemShut {NoStop}%
\bibitem [{\citenamefont {{Adame}}\ \emph {et~al.}(2026)\citenamefont {{Adame}}, \citenamefont {{Avila}}, \citenamefont {{Gonzalez-Perez}}, \citenamefont {{Hahn}}, \citenamefont {{Yepes}},\ and\ \citenamefont {{Manera}}}]{2026A&A...705A..20A}%
  \BibitemOpen
  \bibfield  {author} {\bibinfo {author} {\bibfnamefont {A.~G.}\ \bibnamefont {{Adame}}}, \bibinfo {author} {\bibfnamefont {S.}~\bibnamefont {{Avila}}}, \bibinfo {author} {\bibfnamefont {V.}~\bibnamefont {{Gonzalez-Perez}}}, \bibinfo {author} {\bibfnamefont {O.}~\bibnamefont {{Hahn}}}, \bibinfo {author} {\bibfnamefont {G.}~\bibnamefont {{Yepes}}},\ and\ \bibinfo {author} {\bibfnamefont {M.}~\bibnamefont {{Manera}}},\ }\href {https://doi.org/10.1051/0004-6361/202556174} {\bibfield  {journal} {\bibinfo  {journal} {Astronomy \& Astrophysics}\ }\textbf {\bibinfo {volume} {705}},\ \bibinfo {eid} {A20} (\bibinfo {year} {2026})},\ \Eprint {https://arxiv.org/abs/2506.06200} {arXiv:2506.06200 [astro-ph.CO]} \BibitemShut {NoStop}%
\bibitem [{\citenamefont {{Scoccimarro}}\ \emph {et~al.}(2012)\citenamefont {{Scoccimarro}}, \citenamefont {{Hui}}, \citenamefont {{Manera}},\ and\ \citenamefont {{Chan}}}]{2012PhRvD..85h3002S}%
  \BibitemOpen
  \bibfield  {author} {\bibinfo {author} {\bibfnamefont {R.}~\bibnamefont {{Scoccimarro}}}, \bibinfo {author} {\bibfnamefont {L.}~\bibnamefont {{Hui}}}, \bibinfo {author} {\bibfnamefont {M.}~\bibnamefont {{Manera}}},\ and\ \bibinfo {author} {\bibfnamefont {K.~C.}\ \bibnamefont {{Chan}}},\ }\href {https://doi.org/10.1103/PhysRevD.85.083002} {\bibfield  {journal} {\bibinfo  {journal} {Physical Review D}\ }\textbf {\bibinfo {volume} {85}},\ \bibinfo {eid} {083002} (\bibinfo {year} {2012})},\ \Eprint {https://arxiv.org/abs/1108.5512} {arXiv:1108.5512 [astro-ph.CO]} \BibitemShut {NoStop}%
\bibitem [{\citenamefont {{Balaguera-Antol{\'\i}nez}}\ \emph {et~al.}(2023)\citenamefont {{Balaguera-Antol{\'\i}nez}}, \citenamefont {{Kitaura}}, \citenamefont {{Alam}}, \citenamefont {{Chuang}}, \citenamefont {{Yu}}, \citenamefont {{Favole}}, \citenamefont {{Sinigaglia}}, \citenamefont {{Zhao}}, \citenamefont {{Brooks}}, \citenamefont {{de la Macorra}}, \citenamefont {{Font-Ribera}}, \citenamefont {{Gontcho A Gontcho}}, \citenamefont {{Honscheid}}, \citenamefont {{Kehoe}}, \citenamefont {{Meisner}}, \citenamefont {{Miquel}}, \citenamefont {{Tarl{\'e}}}, \citenamefont {{Vargas-Maga{\~n}a}},\ and\ \citenamefont {{Zhou}}}]{Balaguera_2023}%
  \BibitemOpen
  \bibfield  {author} {\bibinfo {author} {\bibfnamefont {A.}~\bibnamefont {{Balaguera-Antol{\'\i}nez}}}, \bibinfo {author} {\bibfnamefont {F.-S.}\ \bibnamefont {{Kitaura}}}, \bibinfo {author} {\bibfnamefont {S.}~\bibnamefont {{Alam}}}, \bibinfo {author} {\bibfnamefont {C.-H.}\ \bibnamefont {{Chuang}}}, \bibinfo {author} {\bibfnamefont {Y.}~\bibnamefont {{Yu}}}, \bibinfo {author} {\bibfnamefont {G.}~\bibnamefont {{Favole}}}, \bibinfo {author} {\bibfnamefont {F.}~\bibnamefont {{Sinigaglia}}}, \bibinfo {author} {\bibfnamefont {C.}~\bibnamefont {{Zhao}}}, \bibinfo {author} {\bibfnamefont {D.}~\bibnamefont {{Brooks}}}, \bibinfo {author} {\bibfnamefont {A.}~\bibnamefont {{de la Macorra}}}, \bibinfo {author} {\bibfnamefont {A.}~\bibnamefont {{Font-Ribera}}}, \bibinfo {author} {\bibfnamefont {S.}~\bibnamefont {{Gontcho A Gontcho}}}, \bibinfo {author} {\bibfnamefont {K.}~\bibnamefont {{Honscheid}}}, \bibinfo {author} {\bibfnamefont {R.}~\bibnamefont {{Kehoe}}}, \bibinfo {author} {\bibfnamefont {A.}~\bibnamefont
  {{Meisner}}}, \bibinfo {author} {\bibfnamefont {R.}~\bibnamefont {{Miquel}}}, \bibinfo {author} {\bibfnamefont {G.}~\bibnamefont {{Tarl{\'e}}}}, \bibinfo {author} {\bibfnamefont {M.}~\bibnamefont {{Vargas-Maga{\~n}a}}},\ and\ \bibinfo {author} {\bibfnamefont {Z.}~\bibnamefont {{Zhou}}},\ }\href {https://doi.org/10.1051/0004-6361/202245618} {\bibfield  {journal} {\bibinfo  {journal} {Astronomy \& Astrophysics}\ }\textbf {\bibinfo {volume} {673}},\ \bibinfo {eid} {A130} (\bibinfo {year} {2023})},\ \Eprint {https://arxiv.org/abs/2211.10640} {arXiv:2211.10640 [astro-ph.CO]} \BibitemShut {NoStop}%
\end{thebibliography}%

\end{document}